\documentclass[conference]{IEEEtran}
\usepackage{amssymb}
\usepackage{times,amsmath,epsfig,float}
\usepackage{subfigure}
\usepackage{graphicx}
\usepackage{caption2}
\begin{document}
%
\title{Joint Power Adjustment and Receiver Design for Distributed Space-Time Coded in Cooperative MIMO Systems}
%
%
%

\author{\IEEEauthorblockN{Tong Peng, Rodrigo C. de Lamare}
\IEEEauthorblockA{Communications Reasearch Group, Department of Electronics\\University of York, York YO10 5DD, UK\\
Email: tp525@ohm.york.ac.uk; rcdl500@ohm.york.ac.uk}
\and
\IEEEauthorblockN{Anke Schmeink}
\IEEEauthorblockA{UMIC Research Centre\\RWTH Aachen University, D-52056 Aachen, Germany\\
Email: schmeink@umic.rwth-aachen.de}}

\maketitle

\IEEEpeerreviewmaketitle

\begin{abstract}
In this paper, a joint power allocation algorithm with minimum mean-squared error (MMSE) receiver for a cooperative Multiple-Input and Multiple-Output (MIMO) network which employs multiple relays and a Decode-and-Forward (DF) strategy is proposed. A Distributed Space-Time Coding (DSTC) scheme is applied in each relay node. We present a joint constrained optimization algorithm to determine the power allocation parameters and the MMSE receive filter parameter vectors for each transmitted symbol in each link, as well as the channel coefficients matrix. A Stochastic Gradient (SG) algorithm is derived for the calculation of the joint optimization in order to release the receiver from the massive calculation complexity for the MMSE receive filter and power allocation parameters. The simulation results indicate that the proposed algorithm obtains gains compared to the equal power allocation system.

\end{abstract}

\section{Introduction}
 MIMO wireless communication systems employ multiple collocated antennas in both source and destination node in order to obtain the diversity gain and combat multi-path fading. The different methods of STC schemes, which can provide a higher diversity gain and coding gain compared to un-coded systems, are also utilized in MIMO wireless systems. However, it is often impractical to apply MIMO in mobile communication systems due to the high cost and the size of mobile terminals. Cooperative MIMO systems, which employ multiple relay nodes between the source and destination node as the antenna array, apply distributed diversity gain in wireless communication systems \cite{J. N. Laneman2004}. Among the links between the relay nodes and destination nodes, cooperation strategies, such as Amplify-and-Forward (AF), Decode-and-Forward (DF), and Compress-and-Forward (CF) \cite{J. N. Laneman2004} and various DSTC schemes in \cite{J. N. Laneman2003}, \cite{Yiu S.} and \cite{RC De Lamare} can be employed.

Recent contributions in the cooperative communications area lie in the power control problem in transmitters using the AF strategy \cite{O. Seong-Jun}-\cite{Farhadi G.}. A central node which controls the transmission power for each link is employed in \cite{O. Seong-Jun}. Although the central control power allocation can improve the performance significantly, the complexity of the calculation increases with the size of the system. The works on the power allocation problem for the DF strategy measuring the outage probability in each relay node with single antenna and determining the power for each link between the relay nodes and destination node, have been reported in \cite{Min Chen}-\cite{Yindi Jing}. The diversity gain is sacrificed by the utilization of a single antenna in relay nodes.

In this paper, we propose an adaptive power allocation algorithm with linear MMSE receiver for cooperative MIMO systems employing multiple relay nodes with multiple antennas to achieve a DF cooperating strategy. The power allocation parameters and the linear MMSE receive filter parameter vectors can be determined and fad back to each transmission node through a feedback channel that is error free and delay free. The joint estimation algorithm for power allocation parameters in each link and the MMSE receive filter for each symbol is derived. By utilization of an SG algorithm from \cite{S. Haykin}, the complexity of the calculation will be decreased compared with the MMSE-based expressions that require matrix inversions. Also the channel estimation is done by using the SG algorithm before the computation of the power allocation parameters.

The paper is organized as follows. Section II provides the multi-hop cooperative MIMO system with multiple relays applying the DF strategy and DSTC scheme. Section III describes the constrained power allocation problem and linear MMSE detection method, and in Section IV, the proposed iterative SG algorithm is derived. Section V focus on the results of the simulations and Section VI leads to the conclusion.
\begin{figure}
\centering
  \includegraphics[width=3.0in]{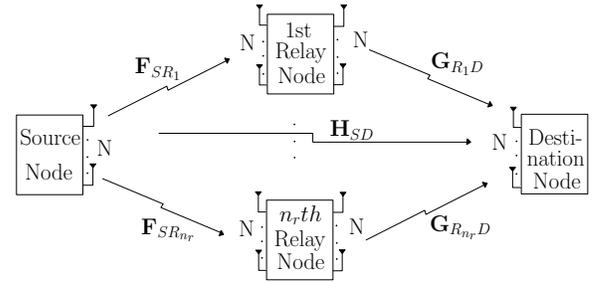}\\
  \caption{Cooperative MIMO System Model with $n_r$ Relay Nodes}\label{1}
\end{figure}

\section{Cooperative System Model}

The communication system under consideration, ${\delta}$ shown in
Fig. 1, is a MIMO communication system transmitting through a
multipath channel from the source node to the destination node.
There are $n_r$  relay nodes, applying a Decode-and-Forward (DF)
scheme as well as space-time coding (STC), between the source and
the destination node, and $N$ antennas at each node for transmitting
and receiving. A multi-hop communication system can be achieved by
broadcasting symbols from the source to $n_r-th$ relay nodes as well
as to the destination node in the first phase, followed by
transmitting the detected and re-encoded symbols from each relay
node to the destination node in the other phases. We consider only
one user at the source node in our system that has $N$ Spatial
Multiplexing (SM)-organized data symbols packed in each packet. The
received symbols at each relay and the destination node, denoted as
${\boldsymbol r}_{{SR}_{n_r}}$ and ${\boldsymbol r}_{SD}$
respectively, are detected by a linear MMSE receive filter at each
receiving node. We assume that the synchronization in each node is
perfect. The transmission between the source node and the $k-th$
relay node, and the destination node can be described as follows
\begin{equation}
{\boldsymbol r}_{{SR}_{k}}[i]  = {\boldsymbol F}_{k}[i]{\boldsymbol A}_{SR_{k}}[i]{\boldsymbol s}[i] + {\boldsymbol n}_{{SR}_{k}}[i],
\end{equation}
\begin{equation}
{\boldsymbol r}_{SD}[i] = {\boldsymbol H}[i]{\boldsymbol A}_{SD}[i]{\boldsymbol s}[i] + {\boldsymbol n}_{SD}[i],
\end{equation}
\begin{equation*}
i = 1,2,~...~,N, ~~k = 1,2,~...~ n_{r}
\end{equation*}
where the $N \times 1$ vector ${\boldsymbol n}_{{SR}_{k}}[i]$ and ${\boldsymbol n}_{SD}[i]$ denote the zero mean complex circular symmetric Additive White Gaussian Noise (AWGN) vector generated in each relay and the destination node with the variance of $\sigma^{2}$. The transmitted symbol vector ${\boldsymbol s}[i] = [s_{1}[i], s_{2}[i], ... , s_{N}[i]]$ contains $N$ parameters, and has a covariance matrix $E\big[ {\boldsymbol s}[i]{\boldsymbol s}^{H}[i]\big] = \sigma_{s}^{2}{\boldsymbol I}$, where $E[ \cdot]$ stands for expected value, $(\cdot)^H$ denotes the Hermitian operator, $\sigma_s^2$ is the signal power which we assume to be equal to 1 and ${\boldsymbol I}$ is the identity matrix. ${\boldsymbol F}_k[i]$ and ${\boldsymbol H}[i]$ are the $N \times N$ channel gain matrix between the source node and the $k-th$ relay node, and between the source node and the destination node, respectively. ${\boldsymbol A}_{SD}[i]$ and ${\boldsymbol A}_{SR_{k}}[i]$ are the diagonal $N \times N$ power allocation matrices with complex parameters $\alpha_{SD}[i]$ and $\alpha_{SR_{k}}[i]$ assigned to each symbol vector ${\boldsymbol s}[i]$ transmitted to the $k-th$ relay node and the destination node.

After filtered in each relay node, the detected symbols will be re-encoded by a $N \times T$ distributed space-time coding (DSTC) matrix and assigned a power allocation parameter matrix and then, forwarded to the destination node. Notice that only the relays which can detect the received symbols correctly will forward the encoded symbols to the destination node because the interference between the received symbols after space-time coding will be increased if the encoded symbol vectors are different. Define $n_{\rm reliable}$ with length $L$ to be the relay set which can implement the correct detection. Then the relation between the $l-th$ relay and the destination node can be described as
\begin{equation}
{\boldsymbol R}_{R_{l}D}[i] = \sum_{t=1}^{T}{\boldsymbol G}_{R_{l}D}[i]{\boldsymbol A}_t[i]{\boldsymbol m}_{{R_{l}D}_t}[i] + {\boldsymbol N}_{R_{l}D}[i],
\end{equation}
\begin{equation*}
    l \in n_{reliable}
\end{equation*}
where the $N \times 1$ matrix ${\boldsymbol m}_{{R_{l}D}_t}[i]$ is the $t-th$ column of the DSTC matrix, $l$ is the number of reliable relay nodes which can detect the received symbols correctly, and ${\boldsymbol A}_t[i]$ is the diagonal matrix contains the power allocation parameter assigned to the $t-th$ column of the re-encoded matrix. The $N \times N$ channel gain matrix is denoted by ${\boldsymbol G}_{R_{l}D}[i]$, and the $N \times T$ AWGN matrix ${\boldsymbol N}_{R_{l}D}[i]$ is generated in the destination node with variance $\sigma^2$. The $N \times T$ received symbol matrix ${\boldsymbol R}_{R_{l}D}[i]$ in (3) can be transformed and expressed as a $NT \times 1$ vector ${\boldsymbol r}_{R_{l}D}[i]$ given by
\begin{equation}
{\boldsymbol r}_{R_{l}D}[i] = \sum_{j=1}^{N}{\boldsymbol D}_{j_{R_{l}D}}[i]{\boldsymbol a}_{j_{R_{l}D}}[i]s_{j_{SR_{l}}}[i] + {\boldsymbol n}_{R_{l}D}[i],
\end{equation}
where the $T \times 1$ vector ${\boldsymbol a}_{j_{R_{l}D}}[i] = [\alpha_{1_{R_{l}D}}[i], ... , \alpha_{T_{R_{l}D}}[i]]$ is the power allocation parameter vector assigned for the $j-th$ symbol $s_j[i]$. The diagonal $NT \times N$ matrix ${\boldsymbol D}_{j_{R_{l}D}}[i] = {\rm diag}[{\boldsymbol d}_{j_{1}}[i], {\boldsymbol d}_{j_{2}}[i], ... , {\boldsymbol d}_{j_{T}}[i]]$ stands for the effective channel coefficient matrix combined with the DSTC scheme and the channel matrix ${\boldsymbol G}_{R_{l}D}[i]$. After rewriting ${\boldsymbol r}_{R_{l}D}[i]$ we can consider the received symbol vector in the destination node as $L+1$ parts, one is from the source node and the remaining $L$ are from the reliable relays, and write the received symbol vector for cooperative detection as
\begin{equation}
\begin{aligned}
{\boldsymbol r}[i]  &=
\left[\begin{array}{c} \sum_{j=1}^{N}{\boldsymbol h}_{j}[i]\alpha_{{j}_{SD}}[i]s_j[i] + {\boldsymbol n}_{SD}[i] \\ \sum_{j=1}^{N}{\boldsymbol D}_{j_{R_{1}D}}[i]{\boldsymbol a}_{j_{R_{1}D}}[i]s_{j_{R_{1}D}}[i] + {\boldsymbol n}_{R_{1}D}[i]  \\ \sum_{j=1}^{N}{\boldsymbol D}_{j_{R_{2}D}}[i]{\boldsymbol a}_{j_{R_{2}D}}[i]s_{j_{R_{2}D}}[i] + {\boldsymbol n}_{R_{2}D}[i] \\ . \\ . \\ . \\ \sum_{j=1}^{N}{\boldsymbol D}_{j_{R_{L}D}}[i]{\boldsymbol a}_{j_{R_{L}D}}[i]s_{j_{R_{L}D}}[i] + {\boldsymbol n}_{R_{L}D}[i] \end{array} \right] \\
 & = \sum_{j=1}^{N}{\boldsymbol B}_j[i]{\boldsymbol a}_j[i]s_j[i] + {\boldsymbol n}[i]
\end{aligned}
\end{equation}
where ${\boldsymbol h}_{j}[i]$ is the $j-th$ column of the $N \times N$ channel coefficients matrix ${\boldsymbol H}[i]$ and $s_{j_{R_{l}D}}[i]={\boldsymbol w}^H_{j_{SR_l}}(\sum_{j=1}^{N}{\boldsymbol f}_{j_{SR_{l}}}[i]\alpha_{{j}_{SR_l}}[i]s_j[i] + {\boldsymbol n}_{SR_l}[i])$ is the detected symbol in the $l-th$ relay node, where ${\boldsymbol f}_{j_{SR_{l}}}[i]$ is the $j-th$ column of the $N \times N$ channel coefficients matrix ${\boldsymbol F}_{SR_{l}}[i]$. The $(LT + 1)N \times (LN + 1)$ diagonal matrix ${\boldsymbol B}_j[i] = {\rm diag} [{\boldsymbol h}_j, {\boldsymbol D}_{j_{R_{1}D}}[i],~...~,~ {\boldsymbol D}_{j_{R_{L}D}}[i]]$ contains the channel gain elements of all the links between the reliable relays and the destination node. The $(LN + 1) \times 1$ power allocation parameter vector ${\boldsymbol a}_j[i] = [\alpha_{j_{SD}}[i], {\boldsymbol a}_{j_{R_{l}D}}[i]]^T = [\alpha_{js}[i], {\boldsymbol a}_{j_{R_{1}D}}[i],~...~,~{\boldsymbol a}_{j_{R_{L}D}}[i]]^T$, where ${\boldsymbol a}_{j_{R_{l}D}}[i]$ denotes the power assigned for the DSTC matrix for $l \in n_{\rm reliable}$, containing all the power elements in each link.

\section{Joint linear MMSE receiver design with Power Allocation}
The MMSE receiver design with power allocation for every link between the source and the destination node as well as the reliable relays is derived as follows. If we define a $(LT + 1)N \times 1$ parameter matrix ${\boldsymbol w}_{j}[i] = [{\boldsymbol w}_{1_1}[i],{\boldsymbol w}_{1_2}[i],...,{\boldsymbol w}_{1_{L+1}}[i]]$ for $i = 1, 2, ..., N_K/N$ to determine the $j-th$ symbol $s_j[i]$ and a $\alpha_{j_{R_{l}D}}[i]$ for $j = 1, 2, ..., N$, by using (5), the MMSE problem with power allocation can be described as
\begin{equation}
\begin{aligned}
& [{\boldsymbol w}_{j,opt}[i],\alpha_{j,opt}[i]] = \arg\min_{{\boldsymbol w}_j[i], \alpha_j[i]} E\big [||s_j[i]-{\boldsymbol w}_j^H[i]{\boldsymbol r}[i]||^2] \\
\end{aligned}
\end{equation}
\begin{equation*}
\begin{aligned}
& {\rm subject} ~~ {\rm to} \\
&\sum_{j=1}^{N}\alpha_{j_{SD}}[i]\alpha_{j_{SD}}^*[i]+\sum_{k=1}^{n_r}\sum_{j=1}^{N}\alpha_{j_{SR_k}}[i]\alpha_{j_{SR_k}}^*[i]=P_T,\\
&\sum_{l=1}^{L}\sum_{j=1}^{N}\alpha_{j_{R_{l}D}}[i]\alpha_{j_{R_{l}D}}^*[i] = P_T
\end{aligned}
\end{equation*}
where $P_T$ is the total transmit power for the source node and all the reliable relay nodes. The joint linear MMSE receiver design and power allocation problem adjusts the receive filter ${\boldsymbol w}_{j,opt}[i]$ and the power allocation parameters $\alpha_{j,opt}[i]$ with a power allocation constraint. It can be transformed into an unconstrained optimization problem using Lagrange multipliers [1] which has the following lagrangian
\begin{equation}
\begin{aligned}
\boldsymbol L=&E\big [||s_j[i]-{\boldsymbol w}_j^H[i]{\boldsymbol r}[i]||^2] + \lambda_1(\sum_{j=1}^{N}\alpha_{j_{SD}}[i]\alpha_{j_{SD}}^*[i]\\
&+\sum_{k=1}^{n_r}\sum_{j=1}^{N}\alpha_{j_{SR_k}}[i]\alpha_{j_{SR_k}}^*[i]-P_T)+\\
&\lambda_2(\sum_{l=1}^{L}\sum_{j=1}^{N}\alpha_{j_{R_{l}D}}[i]\alpha_{j_{R_{l}D}}^*[i]-P_T),
\end{aligned}
\end{equation}
From (7), we can see there are three power allocation parameters to be determined, which are power assigned to the link between the source and the destination node $\alpha_{j_{SD}}[i]$, between the source and each relay node $\alpha_{j_{SR_k}}[i]$ and between the $l-th$ reliable relays and the destination node $\alpha_{j_{R_{l}D}}[i]$. By fixing these $\alpha_j[i]$ and taking gradient terms of (7) and equating them to zero we obtain the expression of ${\boldsymbol w}_{j,opt}[i]$ which is
\begin{equation}
{\boldsymbol w}_{j,opt}[i] = {\boldsymbol R}_{w_{j}[i]}^{-1}{\boldsymbol p}_{w_{j}[i]},
\end{equation}
where ${\boldsymbol R}_{w_{j}[i]}$ is the $(LT + 1)N \times (LT + 1)N$ covariance matrix equals to $E\big [{\boldsymbol r}_j[i]{\boldsymbol r}_j^H[i]\big]$ and the $(LT + 1)N \times 1$ cross-correlation vector ${\boldsymbol p}_{w_{j}[i]} = E\big [{\boldsymbol r}_j[i]s_j^H[i]\big]$. In the expression of ${\boldsymbol R}_{w_{j}[i]}$ and ${\boldsymbol p}_{w_{j}[i]}$ there exists the power allocation parameters. The expression of power allocation parameters can be obtained by fixing ${\boldsymbol w}_{j,opt}[i]$ in (7) and taking gradient terms with respect to $\alpha_{j_{SD},opt}[i]$, $\alpha_{j_{R_{l}D},opt}[i]$ and $\alpha_{j_{SR_k},opt}[i]$ and equating them to zero,
\begin{equation}
\alpha_{j_{SD},opt}[i] = {\boldsymbol R}_{\alpha_{j_{SD}}[i]}^{-1}{\boldsymbol P}_{\alpha_{j_{SD}}[i]},
\end{equation}
\begin{equation}
\alpha_{j_{R_{l}D},opt}[i] = {\boldsymbol R}_{\alpha_{j_{R_{l}D}}[i]}^{-1}{\boldsymbol P}_{\alpha_{j_{R_{l}D}}[i]},
\end{equation}
\begin{equation}
\alpha_{j_{SR_k},opt}[i] = {\boldsymbol R}_{\alpha_{j_{SR_k}}[i]}^{-1}{\boldsymbol P}_{\alpha_{j_{SR_k}}[i]},
\end{equation}
where the covariance matrices are
\begin{equation*}
{\boldsymbol R}_{\alpha_{j_{SD}}[i]} = E\big [{\boldsymbol w}_{j_1}^H[i]{\boldsymbol h}_j[i]s_j[i]s_j^*[i]{\boldsymbol h}_j[i]^H{\boldsymbol w}_{j_1}[i]\big]+\lambda_1
\end{equation*}
\begin{equation*}
\begin{aligned}
{\boldsymbol R}_{\alpha_{j_{R_{l}D}}[i]} = &E\big [{\boldsymbol w}_{j_l}^H[i]{\boldsymbol d}_{j_{R_{l}D}}[i]s_{j}[i]s^*_{j}[i]{\boldsymbol d}^H_{j_{R_{l}D}}[i]{\boldsymbol w}_{j_l}[i]\big ]+\lambda_2
\end{aligned}
\end{equation*}
\begin{equation*}
\begin{aligned}
{\boldsymbol R}_{\alpha_{j_{SR_k}}[i]} = &E\big [{\boldsymbol w}_{j_l}^H[i]({\boldsymbol D}_{j_{R_{l}D}}[i]{\boldsymbol a}_{j_{R_{l}D}}({\boldsymbol w}_{j_{SR_{k}}}^H[i]{\boldsymbol f}_{j_{SR_k}}) s_j[i])\\
&({\boldsymbol D}_{j_{R_{l}D}}[i]{\boldsymbol a}_{j_{R_{l}D}}({\boldsymbol w}_{j_{SR_{k}}}^H[i]{\boldsymbol f}_{j_{SR_k}}) s_j[i])^H{\boldsymbol w}_{j_l}[i]\big ]\\
&+\lambda_1
\end{aligned}
\end{equation*}
and the cross-correlation vectors can be calculated as
\begin{equation*}
\begin{aligned}
{\boldsymbol P}_{\alpha_{j_{SD}}[i]} =& E\big [{\boldsymbol h}^H_j[i]{\boldsymbol w}_{j_1}[i]s_j[i]s_j^*[i]\big ]\\
{\boldsymbol P}_{\alpha_{j_{R_{l}D}}[i]} =& E\big [{\boldsymbol d}^H_{j_{R_{l}D}}[i]{\boldsymbol w}_{j_l}[i]s_j[i]s_j^*[i]\big ]\\
{\boldsymbol P}_{\alpha_{j_{SR_{k}}}[i]} =& E\big [({\boldsymbol w}_{j_l}^H[i]{\boldsymbol D}_{j_{R_{l}D}}[i]{\boldsymbol a}_{j_{R_{l}D}}\\
&({\boldsymbol w}_{j_{SR_{k}}}^H[i]{\boldsymbol f}_{j_{SR_k}}) )^Hs^*_j[i]s_j[i]\big ]
\end{aligned}
\end{equation*}
where ${\boldsymbol w}_{j_l}[i]$ is the MMSE receive filter parameter vector for the $l-th$ receiver symbol vector ${\boldsymbol r}_l[i]$.

The expression of the MMSE detection vector and the power allocation elements depend on each other as well as the effective channel matrix ${\boldsymbol D}_j[i]$, and should be determined by iterating with initial values and the Lagrange multiplier $\lambda_1$ and $\lambda_2$ to obtain the result.

\section{Adaptive Estimation Algorithm for MMSE Design with Power Allocation}
The formulas in the previous section describe the method to calculate the MMSE detection vector ${\boldsymbol w}_j[i]$ and the power allocation parameter $\alpha_j[i]$ for each transmit symbol $s_j[i]$, which require matrix inversions with high complexity as well as channel estimation in the iteration calculation. In this section, an adaptive estimation algorithm based on an SG algorithm will be presented to determine the MMSE receive filter, the power allocation parameters and the effective channel matrix without the inversion calculation.

\subsection{Adaptive SG Estimation for MMSE Receive Filter and Power Allocation}
In this subsection, we will present the adaptive SG estimation algorithm for the MMSE receive filter ${\boldsymbol w}_j[i]$ and the power allocation parameter $\alpha_j[i]$. The problem is described in formula (6) and by using the Lagrange multiplier method [1] we can obtain the expression (7) which indicates the parameters depend on each other. As a result, we will develop a SG joint estimation algorithm with low complexity calculation to solve the problem in (7).

Considering the Lagrangian function in (7) and computing the instantaneous gradient terms of it with respect to ${\boldsymbol w}_j[i]$ and $\alpha_j[i]$, respectively, lead us to the equations (12)-(15),
\begin{table*}
\begin{equation}
\nabla{\boldsymbol L}_{{\boldsymbol w}_j^*[i]}  =  -(\sum_{j=1}^{N}{\boldsymbol B}_j[i]{\boldsymbol a}_j[i]s_j[i] + {\boldsymbol n}[i])(s_j[i] - {\boldsymbol w}_j^H[i](\sum_{j=1}^{N}{\boldsymbol B}_j[i]{\boldsymbol a}_j[i]s_j[i] + {\boldsymbol n}[i]))^*  =  -{\boldsymbol r}[i]e_j^*[i],
\end{equation}
\begin{equation}
\nabla{\boldsymbol L}_{\alpha_{j_{SD}}^*[i]} = -s_j^*[i]{\boldsymbol h}_j^H[i]{\boldsymbol w}_{j_1}[i](s_j[i]- {\boldsymbol w}_j^H[i](\sum_{j=1}^{N}{\boldsymbol B}_j[i]{\boldsymbol a}_j[i]s_j[i] + {\boldsymbol n}[i])) + \lambda_1{\alpha_{j_{SD}}[i]} =  -s_j^*[i]{\boldsymbol h}_j^H[i]{\boldsymbol w}_{j_1}[i]e_j[i] + \lambda_1{\alpha_{j_{SD}}[i]}
\end{equation}
\begin{equation}
\nabla{\boldsymbol L}_{\alpha_{j_{R_{l}D}}^*[i]} = -s_j^*[i]{\boldsymbol d}_j^H[i]{\boldsymbol w}_{j_l}[i](s_j[i]-{\boldsymbol w}_j^H[i](\sum_{j=1}^{N}{\boldsymbol B}_j[i]{\boldsymbol a}_j[i]s_j[i] + {\boldsymbol n}[i]))+\lambda_2{\alpha_{j_{R_{l}D}}[i]} =  -s_j^*[i]{\boldsymbol d}_j^H[i]{\boldsymbol w}^H_{j_l}[i]e_j[i] + \lambda_2{\alpha_{j_{R_{l}D}}[i]}
\end{equation}
\begin{equation}
\begin{aligned}
\nabla{\boldsymbol L}_{\alpha_{j_{SR_{k}}}^*[i]} &=  -s_j^*[i]({\boldsymbol w}^H_{j_l}[i]{\boldsymbol D}_{j_{R_{l}D}}[i]{\boldsymbol a}_{j_{R_{l}D}}[i]{\boldsymbol w}^H_{j_{SR{l}}}[i]{\boldsymbol f}_{j_{SR_{l}}}[i])^H(s_j[i]-{\boldsymbol w}_j^H[i](\sum_{j=1}^{N}{\boldsymbol B}_j[i]{\boldsymbol a}_j[i]s_j[i] + {\boldsymbol n}[i])) +\lambda_2{\alpha_{j_{R_{l}D}}[i]} \\ & = -s_j^*[i]({\boldsymbol w}^H_{j_l}[i]{\boldsymbol D}_{j_{R_{l}D}}[i]{\boldsymbol a}_{j_{R_{l}D}}[i]{\boldsymbol w}^H_{j_{SR{l}}}[i]{\boldsymbol f}_{j_{SR_{l}}}[i])^He_j[i] + \lambda_2{\alpha_{j_{R_{l}D}}[i]}
\end{aligned}
\end{equation}
\rule{18cm}{1pt}
\end{table*}
where ${\boldsymbol d}_j[i]$ is the $j-th$ column of the effective channel matrix ${\boldsymbol D}_{j_{R_{l}D}}[i]$ with dimension $TN \times 1$, and $e_j[i]$ is the error symbol, which indicates the distance between the transmitted symbol and the detected symbol, calculated by ${\boldsymbol s}_j[i] - {\boldsymbol w}_j^H[i]{\boldsymbol r}[i] = s_j[i]-{\boldsymbol w}_j^H[i](\sum_{j=1}^{N}{\boldsymbol B}_j[i]{\boldsymbol a}_j[i]s_j[i] + {\boldsymbol n}[i])$, and $(\cdot)^*$ denotes the conjugation. Notice that we need to determine the MMSE receive filter ${\boldsymbol w}^H_{j_{SR_{k}}}[i]$ at each relay node in order to determine if the $k-th$ relay can detect the received symbols correctly and then forward it to the destination node. We can calculate the instantaneous gradient terms of ${\rm{\boldsymbol  L}}$ with respect to ${\boldsymbol w}_{j_{SR_{k}}}[i]$ to obtain
\begin{equation}
\begin{aligned}
 \nabla{\boldsymbol L}_{{\boldsymbol w}_{j_{SR_{k}}}^*[i]}  = &  -{\boldsymbol w}^H_{j_l}[i]{\boldsymbol D}_{j_{R_{l}D}}[i]{\boldsymbol a}_{j_{R_{l}D}}[i]({\boldsymbol F}_{k}[i]{\boldsymbol A}_{SR_{k}}[i]{\boldsymbol s}[i]\\ & + {\boldsymbol n}_{{SR}_{k}}[i])(s_j[i]- {\boldsymbol w}_j^H[i](\sum_{j=1}^{N}{\boldsymbol B}_j[i]{\boldsymbol a}_j[i]s_j[i] \\ & + {\boldsymbol n}[i]))^* \\
 = & -{\boldsymbol w}^H_{j_l}[i]{\boldsymbol D}_{j_{R_{l}D}}[i]{\boldsymbol a}_{j_{R_{l}D}}[i]{\boldsymbol r}_{{SR}_{k}}[i]e_j^*[i],
\end{aligned}
\end{equation}
We can devise an adaptive SG estimation algorithm by using the instantaneous gradient terms of the Lagrangian which were previously derived with the SG descent rules [1] as
\begin{equation}
{\boldsymbol w}_j[i + 1] = {\boldsymbol w}_j[i] - \mu\nabla{\boldsymbol L}_{{\boldsymbol w}_j^*[i]}
\end{equation}
\begin{equation}
{\boldsymbol w}_{j_{SR{k}}}[i + 1] = {\boldsymbol w}_{j_{SR_{k}}}[i] - \mu\nabla{\boldsymbol L}_{{\boldsymbol w}_{j_{SR_{k}}}^*[i]}
\end{equation}
\begin{equation}
\begin{aligned}
\alpha_{j_{SD}}[i + 1] & = \alpha_{j_{SD}}[i] - \gamma\nabla{\boldsymbol L}_{\alpha_{j_{SD}}^*[i]} \\
\end{aligned}
\end{equation}
\begin{equation}
\begin{aligned}
\alpha_{j_{R_{l}D}}[i + 1] & = \alpha_{j_{R_{l}D}}[i] - \gamma\nabla{\boldsymbol L}_{\alpha_{j_{R_{l}D}}^*[i]} \\
\end{aligned}
\end{equation}
\begin{equation}
\begin{aligned}
\alpha_{j_{SR_{l}}}[i + 1] & = \alpha_{j_{SR_{l}}}[i] - \gamma\nabla{\boldsymbol L}_{\alpha_{j_{SR_{l}}}^*[i]} \\
\end{aligned}
\end{equation},
where $\mu$ and $\gamma$ are the step sizes of the recursions for the estimation of MMSE parameter vectors which have to be determined before the estimation. The complexity of calculating ${\boldsymbol w}_j[i]$ is $(\boldsymbol O(N))$ and $(\boldsymbol O(N))$ for calculating $\alpha_{j_{SD}}[i]$ and $\alpha_{j_{R_{l}D}}[i]$, and $(\boldsymbol O(TN))$ for $\alpha_{j_{SR_{l}}}[i]$, which are much less than that of the algorithm we described in Section III. As mentioned in Section II, all the MMSE receiver filters and power allocation matrices will be transmitted back to the relay nodes via a feedback channel which is assumed to be error-free in the simulation; however, in practical situation the errors at each relay node should be considered due to the property of broadcasting and the diversification of the feedback channels with time changes.

\subsection{Adaptive SG Channel Estimation}
In this subsection we will derive an adaptive SG algorithm for estimating the effective channel matrix ${\boldsymbol B}_j[i]$. The channel estimation can be described as an optimization problem
\begin{equation}
{\boldsymbol B}_j[i] = \arg \min_{{\boldsymbol B}_j[i]}{E\big [||{\boldsymbol r}[i] - {\boldsymbol B}_j[i]{\boldsymbol a}_j[i]s_j[i]||^2]}
\end{equation}
Define the received symbol vector ${\boldsymbol r}[i] = [{\boldsymbol r}_1[i],...,{\boldsymbol r}_{L+1}[i]]$. The optimization problem described in (22) can be divided into three parts, which correspond to individually computing ${\boldsymbol H}[i]$, ${\boldsymbol F}_{SR_{k}}[i]$ and ${\boldsymbol D}_{j_{R_{l}D}}[i]$. We then derive an SG algorithm by calculating the cost function ${\rm C}_{\boldsymbol H}$, ${\rm C}_{\boldsymbol F}$ and ${\rm C}_{\boldsymbol D}$
\begin{equation}
 C_{\boldsymbol H}  =  E\big [||{\boldsymbol r}_1[i] - {\boldsymbol H}[i]{\boldsymbol A}_{SD}[i]{\boldsymbol s}[i]||^2\big ]
\end{equation}
\begin{equation}
\begin{aligned}
 C_{\boldsymbol F}  = & E\big [||{\boldsymbol r}_l[i] - \sum_{j=1}^{N}{\boldsymbol D}_{j_{R_{l}D}}[i]{\boldsymbol a}_{j_{R_{l}D}}[i]{\boldsymbol w}^H_{j_{SR_{l}}}[i]({\boldsymbol F}_{SR_{l}}[i]\\
 &{\boldsymbol A}_{SR_{l}}[i]{\boldsymbol s}[i] + {\boldsymbol n}_{SR_{l}}[i])||^2\big ] \\
\end{aligned}
\end{equation}
\begin{equation}
 C_{\boldsymbol D}  =  E\big [||{\boldsymbol r}_l[i] - \sum_{j=1}^{N}{\boldsymbol D}_{j_{R_{l}D}}[i]{\boldsymbol a}_{j_{R_{l}D}}[i]s_{j_{SR_{l}}}[i]||^2\big ]
\end{equation}
and then by taking instantaneous gradient terms of ${\rm C}_{\boldsymbol H}$, ${\rm C}_{\boldsymbol F}$ and ${\rm C}_{\boldsymbol D}$ with respect to ${\boldsymbol H}[i]$, ${\boldsymbol F}[i]$ and the $jth$ column of effective channel vector ${\boldsymbol D}_{j_{R_{l}D}}[i]$
\begin{equation}
\nabla C_{{\boldsymbol H}^*[i]} = -{\boldsymbol A}_{SD}[i]{\boldsymbol s}[i]({\boldsymbol r}_1[i] - {\boldsymbol H}[i]{\boldsymbol A}_{SD}[i]{\boldsymbol s}[i]),
\end{equation}
\begin{equation}
\begin{aligned}
\nabla C_{{\boldsymbol F}^*[i]} =& -\sum_{j=1}^{N}{\boldsymbol D}_{j_{R_{l}D}}[i]{\boldsymbol a}_{j_{R_{l}D}}[i] {\boldsymbol w}^H_{j_{SR_{l}}}[i]({\boldsymbol A}_{SR_{l}}[i]{\boldsymbol s}[i])\\
&({\boldsymbol r}_l[i] - \sum_{j=1}^{N}{\boldsymbol D}_{j_{R_{l}D}}[i]{\boldsymbol a}_{j_{R_{l}D}}[i]{\boldsymbol w}^H_{j_{SR_{l}}}[i]{\boldsymbol r}_{SR_{l}}),
\end{aligned}
\end{equation}
\begin{equation}
\begin{aligned}
\nabla C_{{\boldsymbol D}^*[i]} &= -{\boldsymbol a}_{j_{R_{l}D}}[i]s_j[i]({\boldsymbol r}_l[i]-\sum_{j=1}^{N}{\boldsymbol D}_{j_{R_{l}D}}[i]{\boldsymbol a}_{j_{R_{l}D}}[i]\\
&s_{j_{SR_{l}}}[i]),
\end{aligned}
\end{equation}
Considering the SG descent rules [1] and the result of gradient terms we can obtain the adaptive SG channel estimation expression which is
\begin{equation}
{\boldsymbol H}[i + 1] = {\boldsymbol H}[i] - \beta\nabla C_{{\boldsymbol H}^*[i]}
\end{equation}
\begin{equation}
{\boldsymbol F}[i + 1] = {\boldsymbol F}[i] - \beta\nabla C_{{\boldsymbol F}^*[i]}
\end{equation}
\begin{equation}
    {\boldsymbol D}_{j_{R_{l}D}}[i + 1] = {\boldsymbol D}_{j_{R_{l}D}}[i] - \beta\nabla C_{{\boldsymbol D}_{j_{R_{l}D}}^*[i]}
\end{equation}
where $\beta$ is the step size of the recursion. The adaptive SG algorithm for effective channel ${\boldsymbol H}[i]$, ${\boldsymbol F}[i]$ and ${\boldsymbol D}_{j_{R_lD}}[i]$ requires the calculation complexity of $(\boldsymbol O(N^2))$, $(\boldsymbol O(TN))$ and $(\boldsymbol O(TN))$ and can determine the channel matrix accurately.

\section{Simulations}
The simulation results are provided in this section to assess the proposed algorithm. The system we considered is a DF cooperative MIMO system with different distributed STC schemes using QPSK modulation in quasi-static block fading channel with Additive White Gaussian Noise (AWGN), as derived in Section II. The bit error ratio (BER) performance of the joint power allocation using a linear MMSE receive filter (JPA-LMF) algorithm and the equal power allocation using a linear MMSE receive filter (EPA-LMF) algorithm with the power constraint employs different number of relay nodes and different STC schemes are compared. In the simulation we define the power constraint $P_T$ as equal to 1, and the noise variance $\sigma^{2}$ for each link is equal to 1 as well.

The proposed JPA-LMF algorithm is compared with the EPA-LMF algorithm using the distributed-Alamouti (D-Alamouti) STBC scheme in \cite{RC De Lamare} with $n_r = 1,2$ relay nodes in Fig. 2. The number of antennas $N=2$ at each node and the cooperative decoding delay at the destination node is $T=2$ time slots. The results illustrate that the performance of EPA-LMF algorithm is close to that of JPA-LMF algorithm when using the same DSTC scheme in lower $E_b/N_0$ circumstance; however, with the $E_b/N_0$ increase, the JPA-LMF algorithm obtains about 5dB of gains compared to the EPA-LMF algorithm using the same DSTC scheme to achieve the identical BER. The performance improvement of the proposed JPA-LMF algorithm is achieved with more relays employed in the system as an increased spatial diversity is provided by the relays.
Fig. 3 illustrates the performance of the proposed JPA-LMF algorithm using different STC schemes under the same $E_b/N_0$ condition and which employs $n_r = 1$ relay node. The cooperative detection is achieved using a linear MMSE receive filter at the destination node. The randomised-Alamouti (R-Alamouti) STBC \cite{Birsen Sirkeci-Mergen} is employed in the simulation and the number of antennas is set to $N=2$ in each node. The simulation results shown in Fig. 3 indicate that the proposed JPA-LMF algorithm using R-Alamouti STBC obtains about 5dB of gain compared to the EPA-LMF algorithm using the same DSTC scheme. There exists 3.5dB of gain of the simulation result for the R-Alamouti STBC utilizing the JPA-LMF algorithm compared to that for the D-Alamouti STBC utilizing the same algorithm to achieve an identical BER as illustrated in Fig. 3.

\section{Conclusion}

We have proposed a joint power allocation and receiver design algorithm using a linear MMSE receive filter with the power constraint between the source node and the relay nodes, and relay nodes and the destination node. A joint iterative estimation algorithm for computing the power allocation parameter vector and linear MMSE receive filter has been derived. The simulation results illustrate the advantage of the proposed power allocation algorithm by comparing it with the equal power algorithm. The proposed algorithm can be utilized with different distributed STC schemes using the DF strategy, can be extended to the AF cooperation protocols and non-linear receivers\cite{RC De Lamare2008}.



\ifCLASSOPTIONcaptionsoff
  \newpage
\fi

\end{document}